\documentclass[aip,jcp,floatfix,reprint]{revtex4-1}

\usepackage{graphicx}
\usepackage{amsmath}
\usepackage{amssymb}
\usepackage{amsfonts}
\usepackage{color}

\usepackage{times}

\graphicspath{{./FIGS/}}

\begin{document}

\title{The effect of topology on the structure and free energy landscape of
DNA kissing complexes}

\author{Flavio Romano}
%\affiliation{Physical and Theoretical Chemistry Laboratory, 
% Department of Chemistry, University of Oxford, South Parks Road, 
% Oxford, OX1 3QZ, United Kingdom}
\author{Alex Hudson}
\thanks{Present address: Department of Chemistry, University of California,
Berkeley, California 94720, USA}
%\affiliation{Physical and Theoretical Chemistry Laboratory, 
% Department of Chemistry, University of Oxford, South Parks Road, 
% Oxford, OX1 3QZ, United Kingdom}
\author{Jonathan P.~K.~Doye}
\thanks{Author for correspondence}
\affiliation{Physical and Theoretical Chemistry Laboratory, 
 Department of Chemistry, University of Oxford, South Parks Road, 
 Oxford, OX1 3QZ, United Kingdom}
\author{Thomas E.~Ouldridge}
%\affiliation{Rudolf Peierls Centre for Theoretical Physics, 
% University of Oxford, 1 Keble Road, Oxford, OX1 3NP, United Kingdom}
\author{Ard A.~Louis}
\affiliation{Rudolf Peierls Centre for Theoretical Physics, 
 University of Oxford, 1 Keble Road, Oxford, OX1 3NP, United Kingdom}

\date{\today}

\begin{abstract}
We use a recently developed coarse-grained model for DNA to study 
kissing complexes formed by hybridization of complementary hairpin loops.
  The binding of the loops is topologically constrained because their linking
number must remain constant. By studying systems with  linking numbers 
-1, 0 or 1 we show that  the average number of interstrand base pairs is larger when the topology is more favourable for 
 the right-handed wrapping of strands around each other. The thermodynamic stability of the kissing complex also decreases when the linking number changes from -1 to 0 to 1.  The structures of the kissing complexes typically involve
two intermolecular helices that coaxially stack with the hairpin stems at 
a parallel four-way junction.
  \end{abstract}

\maketitle

\section{Introduction}
%DNA molecules have the unusual property of being able to store information in the sequence of bases attached to their sugar and phosphate backbone. 
Not only is DNA the genetic information carrier of  life, but, given the degree of control achieved in the chemistry of
DNA\cite{Car09a} --- molecule synthesis is fast, reliable and relatively
cheap --- these information-rich building blocks can be exploited to reliably
self-assemble two- and three-dimensional 
structures\cite{See1982a,Seeman2003,Seeman10} and 
to build functional nanodevices.\cite{Bath2007}

Hairpins probably represent the simplest structure that DNA can form
besides the standard double helix. These are secondary structure motifs
formed by single-stranded DNA molecules that have complementary regions
that self-hybridize. The intramolecular double helix formed from the
self-complementary sections is known as stem or neck, while the section
that connects two of the stem ends is called a loop.

%Hairpins are perhaps the simplest and most
%common structures that ssDNA can form, and are present both in biologically
%and technologically relevant systems, either as stable or metastable
%states, and their presence can be desired --- they are used as a
%catalyst/fuel or to create tension in DNA nanotechnology --- or undesired,
%as it is often the case in DNA self-assembly. Hairpin regions are quite
%common in long strands, and are an essential feature of secondary and
%tertiary structures, both for their shape and for the possibility to add
%tension to the strand itself.

In contrast to RNA, which for the most part is single-stranded {\em in
vivo}, so that hairpins are a common structural element,\cite{Varani95} DNA
{\em in vivo} is mostly in its duplex form. Nevertheless, there are
occasions when it is single-stranded, and examples have been identified
where DNA hairpins play a biological role,\cite{Bikard10} including in
replication, transcription and recombination.\cite{GlucksmannKuis96,
Oettinger04} However, hairpin formation can sometimes be an undesired
process, and has been implicated in certain
diseases.\cite{Santhana96,Lam11}

DNA hairpins also play an important role in DNA
nanotechnology\cite{Dirks04,Boi05a,Gre06a,Venkataraman2007,Green2008,Yin2008,Muscat2011}
and can be used as the ``fuel'' to provide the energy to power autonomous
DNA motors,
%\cite{Venkataraman2007,Green2008,Yin2008,Muscat2011},
although they can also be an unwanted secondary structural motif in DNA
designed to be unstructured.\cite{Ouldridge_tweezers_2010}

In this paper we study ``kissing'' complexes that can form when the loops
of two DNA hairpins are complementary and partially hybridize. In
particular, we focus on the interplay between topology and the shape and
stability of these complexes. For example, when two hairpin loops
hybridize, the right-handed wrapping of the DNA strands in the
intermolecular double helix must be compensated by a region where the loops
wrap around each other in the opposite sense. Thus, even when the two loops
are fully complementary, topological effects will restrict the number of
bonds that can be formed.

Kissing loop interactions are also an important RNA tertiary structure
motif~\cite{Tin99a} and play key biological roles in processes such as the
regulatory action of antisense RNAs and the dimerization of viral genomic
RNA.\cite{Bois9} They have therefore been much better characterized for
RNA, both in terms of their structure~\cite{Bindewald08,Lee98a,Reb03a} and
mechanical properties,~\cite{Li06,Li09d} than for DNA. This structural
knowledge has even been exploited in structural RNA nanotechnology, where
kissing loop interactions have also been used as a means to join RNA
components with a well-defined geometry.\cite{Horiya03,Severcan10,Grabow11}
Although these RNA systems provide an interesting comparison, the kissing
loop interactions typically involve shorter sequences of complementary
bases than for the DNA systems we consider here, and so topological effects
are less significant. Interestingly, the NMR solution structure of a DNA
kissing complex has been obtained for sequences analogous to that of a
previously characterized RNA kissing complex.\cite{Barbault02} Although
there are differences in the details of the two structures, they are
generally very similar.

%Two DNA hairpins that have complementary loops can form a so-called kissing
%complex when the loops (partially) hybridise. 
%One of the interesting aspects of kissing complexes is the potential topological effects that can lead to binding being restricted, because as the two hairpin loops hybridize the right-handed wrapping of the DNA strands in the intermolecular double helix must be compensated by a region where the loops wrap around each other in the opposite sense.

The topological effects associated with kissing loop interactions have been 
exploited in DNA nanotechnology, particularly to allow the design of 
autonomous motors.\cite{Turberfield2003,Boi05a,Gre06a,See06a}
One way of driving a DNA nanodevice through a cycle is through the use of 
complementary single-stranded DNA strands as ``fuel'', 
where the first strand is designed to partially hybridize with the device to 
induce a conformational change, and the second then reverses this change by 
displacing it to form a ``waste'' duplex. The first example of such a device
was DNA nanotweezers, where the strands induced the device to open and 
close.\cite{Yurke2000} However, one problem with such a device is the
two strands have to be added sequentially, since, if both are present at
the same time, they will preferentially directly hybridize with each other
rather than with the device. 

%To circumvent this problem, one needs a way of preventing the fuel strands hybridizing except when catalyzed by the device.
One way to circumvent this problem is through the use of fuel strands that can form 
hairpins,\cite{Boi05a,Gre06a} since the topological restriction on 
binding between the hairpin loops will effectively prevent the duplexes from being 
formed, even though the duplexes are more stable. Given that the two strands are complementary, these hairpins naturally form kissing complexes.
%render duplex formation extremely slow
%Even though they are more stable than the hairpins --- 
While the hairpins are unable to open each other's stems by displacement, 
the motor can be designed to be a catalyst for the
hybridization. By having a single-stranded DNA section that both is partially 
complementary to one of the hairpins and has a free end, the motor can
open the hairpin by displacement, unconstrained by topological effects.
A second, and similar, solution
is to prepare the fuel strands complexed to partially complementary 
protective strands that bind to either end of the fuels but leave a loop
region in the middle unhybridized.\cite{Turberfield2003,See06a}
A variety of autonomous motors have been designed based on these 
principles.\cite{Venkataraman2007,Green2008,Yin2008,Muscat2011}

Here, we investigate the system of fully complementary 40 base DNA hairpins
studied by Bois {\it et al.}\cite{Boi05a} using computer simulations of a
nucleotide-level coarse-grained model of
DNA.\cite{Ouldridge_tweezers_2010,Ouldridge_long_2011} This recently
introduced model provides an excellent description of the structural,
thermodynamic and mechanical properties of both single-stranded and duplex
DNA, and has now made it feasible to study the free energy landscapes of
such DNA nanotechnology systems in detail, as previously illustrated for
DNA nanotweezers.\cite{Ouldridge_tweezers_2010} In particular, we focus on
the effects of topology on the free energy landscape for the binding
of the hairpin loops, and how the structure of the resulting kissing
complex reflects these topological constraints. To further illustrate the
role played by the topology, we also consider kissing complex formation in
systems of linked hairpins.

\section{Model and Methods}

\subsection{Model}
We use the coarse-grained DNA model developed by Ouldridge {\em et.
al.}.\cite{Ouldridge_tweezers_2010,Ouldridge_long_2011} In this model, a
DNA strand is described as a polymer of nucleotides that interact via
excluded volume repulsion and anisotropic attractive potentials that mimic
the Watson-Crick base-pairing, stacking, cross stacking and coaxial
stacking. The model has been parameterized to reproduce the structural and
thermodynamic properties of single-stranded and double-stranded DNA
molecules at the high salt concentrations that are typically used in DNA
nanotechnology applications. Since this model is described in detail in
Ref.~\onlinecite{Ouldridge_long_2011} we shall repeat here only the
fundamental ingredients.

Each nucleotide is represented as a rigid body with three interaction
sites, all on the same axis. Although the interaction sites are collinear,
we stress that a nucleotide does not possess cylindrical
symmetry, since the potential also depends on a vector
perpendicular to the nucleotide axis to capture the effects of the
orientation of a base on the interactions.

The potential energy $V$ can be written as:
\begin{equation}
\label{eq:pot}
\begin{split}
V &=\sum_{\rm nn}{(V_{\rm backbone} + V_{\rm stack} + V^{\prime}_{\rm
exc})}\ + \\ &\sum_{\rm other\ pairs}(V_{\rm HB} + V_{\rm cross.~stack} +V_{\rm
coax~stack} + V_{\rm exc})\:.
\end{split}
\end{equation}
The first sum runs over all pairs of nucleotides that are adjacent along a
strand (neighbours in our terminology) and the second sum runs over all
other pairs. The interaction between neighbours consists of a backbone term
that is designed to represent the connectivity of a DNA strand, a stacking
term that is designed to mimic stacking interaction between nucleotides,
and an excluded volume part that prevents nucleotides from overlapping. The
interaction between non-neighbouring pairs consists of four different
terms: (i) a hydrogen-bonding term that mimics directional Watson-Crick
base-pairing; (ii) a cross-stacking term that accounts for stacking
interactions between nucleotides that are second neighbours on different
strands; (iii) a coaxial stacking term that is designed to capture stacking
interactions between non-neighbouring bases; and (iv) an excluded volume
term. The full forms of each of these terms is reported in
Ref.~\onlinecite{Ouldridge_long_2011}, with the exception of the coaxial
stacking term, which is described in Ref.~\onlinecite{Ouldridge11b}. Its
parametrization will be described in detail elsewhere.\cite{Ouldridge12}

Features of the model that are particularly important for the current study
are the relative flexibility of single-stranded DNA and its ability to
describe the thermodynamics of hairpin formation accurately, as well as
hybridization in general. We are also confident of the general robustness
of the model, based on the wide range of DNA systems on which
we\cite{Ouldridge_tweezers_2010,Ouldridge_long_2011,Ouldridge11b} and
others\cite{CDM} have tested the model. These so far include DNA
nanotweezers,\cite{Yurke2000} ``burnt bridges''\cite{Bath2005} and
two-footed\cite{Bath2009} DNA walkers, as well as processes such as DNA
displacement,\cite{Zhang_disp_2009} overstretching\cite{Smith96} and
cruciform formation\cite{Ramreddy11} and the formation of liquid
crystalline phases.\cite{BelliniScience}

However, we should also note that the model does introduce a significant
level of coarse graining and neglects several features of the DNA structure
and interactions. Firstly, all four nucleotides have the same structure and
interaction properties, except for the hydrogen-bonding term, for which
interactions are only allowed between Watson-Crick complementary bases.
Although this approximation of course precludes the study of much of the
sequence dependence of properties and behaviour, it is not a problem when,
as here, we are interested in the generic behaviour of a system. Secondly,
the double helix in our model is symmetrical rather than having different
sizes for the minor and major grooves. Again this is unlikely to be an
issue, unless we are interested in the DNA structure at a quite fine level
of detail. Finally, the interactions have been fitted for a single, fairly
high, salt concentration (namely, 0.5M), where the Debye screening length
is short. This is the regime relevant to most DNA nanotechnology
experiments.

\subsection{Simulation Methods}
Throughout this work, we use Monte Carlo simulations employing the Virtual
Move algorithm (VMMC) introduced by Whitelam and
coworkers.\cite{Whitelam2007, Whitelam2009} The latter is a modification to
the standard (Metropolis) Monte Carlo algorithm specifically designed to
promote the collective diffusion of highly interacting clusters that would
otherwise be suppressed. In our model DNA strands are effectively clusters
of interacting nucleotides and that VMMC significantly speeds up sampling,
particularly when strand diffusion is important, which is the case when
studying hybridization processes.

Because of the presence of large free-energy barriers, we have used
umbrella sampling~\cite{Torrie1977} to accurately sample transitions
between different states. In practice, this is accomplished by adding an
additional term to the system Hamiltonian designed to flatten the free
energy profile along a particular reaction coordinate, and then
subsequently unbiasing the results.\cite{Kumar1992} In the present case,
the natural choice for the reaction coordinate is the number of base pairs
between the two strands. This choice requires a definition of a base
pairing, and we define a pair of nucleotides as base paired if the hydrogen
bonding interaction term between them is at least $0.093$ times the well
depth. Of course this choice is somewhat arbitrary, but changing the
threshold does not significantly alter the results.

\begin{figure}[tb]
\begin{center}
\includegraphics[width=8.5cm]{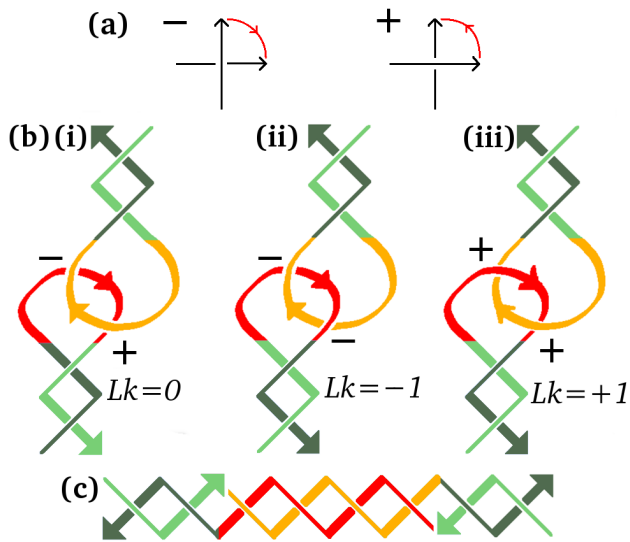}
\end{center}
\caption{\label{fig:schematics}
(a) The definitions that we use for the signs associated with the crossing 
of two curves.
Schematic representations of (b) the three topological
configurations of DNA hairpins and studied in this paper and (c) a control
system with no topological or geometric constraints. The different
topological configurations in (b) are: 
(i) Topologically unlinked, linking number $Lk=0$. 
(ii) Topologically favoured, $Lk=-1$. 
(iii) Topologically frustrated $Lk=+1$. 
}
\end{figure}

\subsection{DNA Sequences}
We have studied the same 40-base nucleotide sequences as in
Ref.~\onlinecite{Boi05a}. The two DNA strands are fully complementary, and so
can form a duplex as well as hairpins with a stem of 10 base
pairs and a loop of 20 bases. 
The sequences of the two strands are, in $5^\prime$ to $3^\prime$ direction:
\begin{center}
\begin{small}
\begin{verbatim}
gcgttgctgc-attttactcttctcccctcg-gcagcaacgc
\end{verbatim}
\end{small}
\end{center}
and
\begin{center}
\begin{small}
\begin{verbatim}
gcgttgctgc-cgaggggagaagagtaaaat-gcagcaacgc
\end{verbatim}
\end{small}
\end{center}
where the hyphens separate stem and loop regions. 
All the results we present are at room temperature, taken as
$T=296.15\,{\rm K}$
($23\,{\rm ^{\circ}C}$). This compares to a hairpin melting temperature of
around $350\,{\rm K}$. Hence, at the temperature we consider, the probability of
spontaneous hairpin opening is effectively zero in our simulations.

\subsection{DNA Topology}
A commonly used property to characterise systems with respect to their
topology is the linking number, $Lk$, a number that describes how two
closed curves are linked in three-dimensional space. Given the projection
of two closed curves onto any plane, a crossing is taken to be positive
(negative) if the upper curve can be superimposed onto the lower by a
counterclockwise (clockwise) rotation (see Fig.~\ref{fig:schematics}(a)).
According to this definition, each crossing in the right-handed helix
formed by dsDNA is negative as the strands in the helix are antiparallel.~\cite{DNAtopology_signs}
The linking number $Lk$ is then defined as
\begin{equation}
Lk = \frac{1}{2}\sum_{i}{c_i}
\end{equation}
where the index $i$ runs over the crossings with $c_i=+1$ for a positive
crossing and $c_i=-1$ for a negative crossing. In more intuitive terms, $Lk$
is the number of times that each curve wraps around the other. For a detailed
discussion of the linking number in the context of nucleic acids we refer the
reader to Ref.~\onlinecite{DNAtopology}.

Although we are considering a system of two unclosed DNA strands, because
the rate of hairpin opening is negligible in our simulations at the
temperature we consider, the hairpin loops can be effectively considered as
closed loops, and so can exhibit topologically different states
(Fig.~\ref{fig:schematics}(b)). The most experimentally relevant state is
that with $Lk=0$, as the likelihood that two hairpin loops would interlink
during their formation process is very low. Since linking number is
conserved, in this state any negative crossings of the two hairpin loops
due to hybridization of the complimentary loops must be compensated by
positive crossings, i.e.\ sections where the loops wind round each other
in the opposite sense. This topological effect will frustrate the
hybridization process.

\section{Results}

\subsection{\label{sec:0}Topologically unlinked complex; $Lk=0$}
\begin{figure}[tb]
\begin{center}
\includegraphics[width=8.5cm]{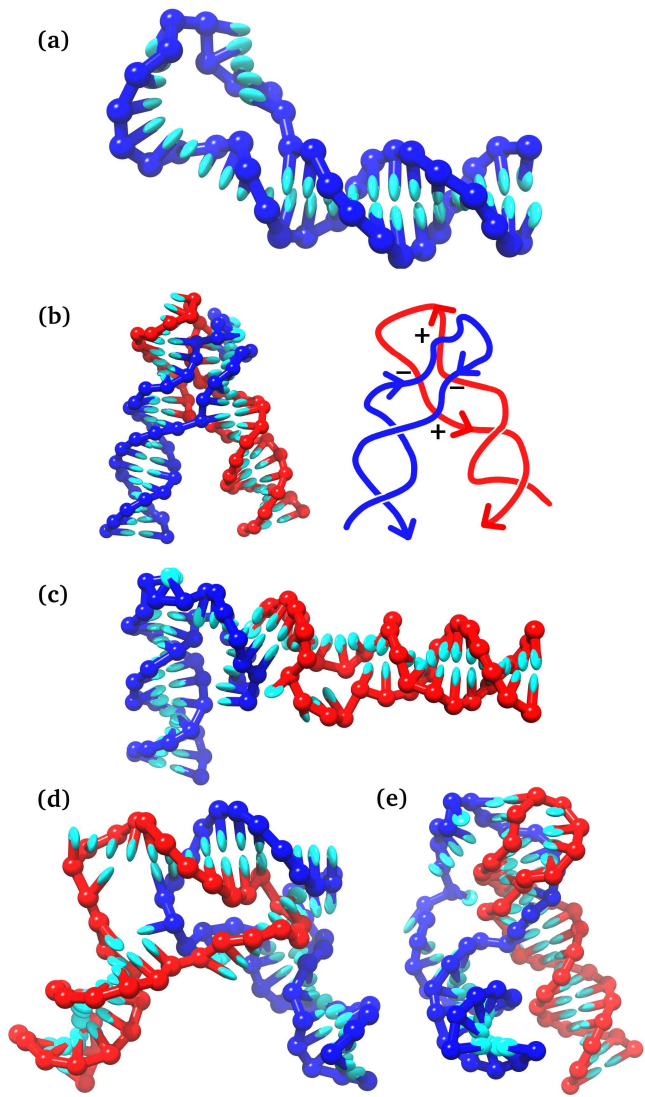}
\end{center}
\caption{\label{fig:struct1}
Typical structures assumed by (a) a single hairpin and (b) the
topologically unlinked ($Lk=0$) kissing complex. Note that in (a) it almost
looks as if the stem is longer than 10 base pairs, because the stacking
tends to propagate beyond the end of the stem at this temperature. For (b)
the chosen structure has 14 interstrand base pairs. To its right is a
topological sketch of the configuration illustrating that the zero linking
number is achieved by balancing positive and negative crossings. In panels
(c)--(e) we show example structures for partially formed complexes with a
total of 2, 6 and 10 base pairs formed between the loops, respectively.
}
\end{figure}

\begin{figure}[tb]
\includegraphics[width=8.5cm]{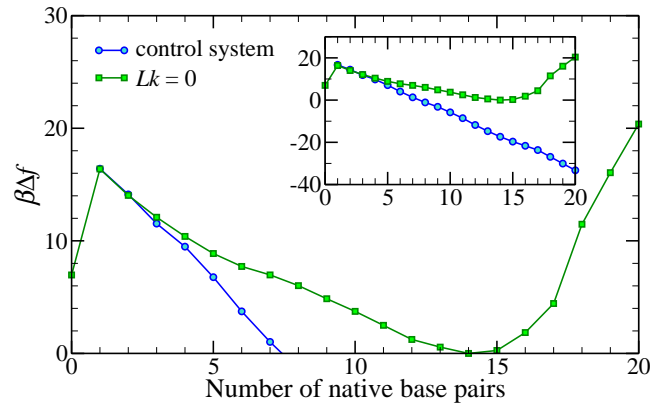}
\caption{\label{fig:fel2}
Free energy profile of two complementary hairpins that are topologically
unlinked (i.e. $Lk=0$) at $T=23\,^\circ{\rm C}$ at a single strand
concentration of $0.336\,$mM (squares). The free energy profile for
hybridization of the control system (Fig.~\ref{fig:schematics}(c)) is also
plotted (circles). In the inset, the full profiles are plotted, showing
the large ($>30\,k_{\rm B}T$) free energy difference between the most
stable states of the kissing complex and the control system. The free
energy profiles have been taken to have the same value when the number of
base pairs is $1$, assuming the same value for the association barrier.
}
\end{figure}

\begin{figure}[tb]
\begin{center}
\includegraphics[width=8.5cm]{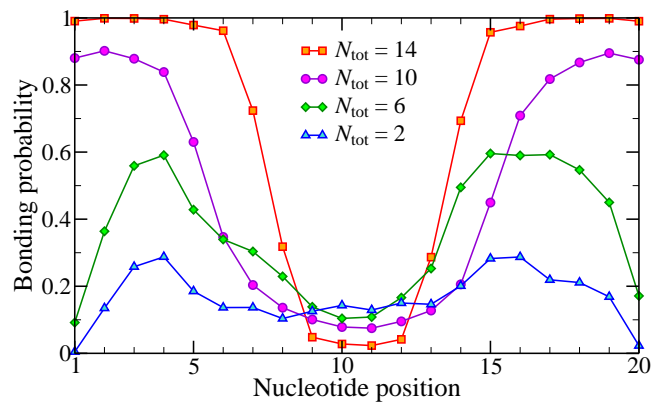}
\end{center}
\caption{\label{fig:dmms}
Bonding probability as a function of nucleotide position in the loop for
different $N_{\rm tot}$, the total number of correct base pairs in the
kissing complex. Note that the probability is not completely symmetric
around the centre of the loop sequence because the propensity to form
non-native base pairs depends on which native base pairs are formed.
}
\end{figure}

Firstly, we consider two hairpins that are topologically unlinked
(Fig.~\ref{fig:schematics}(b)(i)) and are free to bind through the
complementary loop regions. One of the unbound hairpins is illustrated in
Fig.\ \ref{fig:struct1}(a). The room temperature free energy profile for
bonding is reported in Fig.~\ref{fig:fel2}. The jump associated with the
formation of the first bond is due to the loss of translational entropy
associated with hybridization, and is dependent on concentration as well as
temperature. Our data were collected with two strands in a volume of
$4944\,{\rm nm}^3$, i.e.\ $0.336\,$mM. 
%See later in the text for a discussion of how our results relate to bulk
%yields of kissing complexes.

For the hybridization of a duplex in our model, we have previously shown
that, after the barrier for forming the first base pair, there is then a
linear decrease in the free energy as the number of base pairs increases
(aside from a possible small rise at the end due to the fraying of the ends
of the duplex).\cite{Ouldridge_long_2011} The behaviour seen for the
binding of the two hairpin loops is significantly different from this
scenario. After an initial roughly linear decrease (up to about 3--4 base
pairs), the line begins to exhibit some positive curvature reaching a
minimum at 14 base pairs (roughly one and a half helical turns) before
rising steeply. This curvature is a result of the topological constraint
that as the two hairpin loops wind around each other to form a duplex the
linking number must remain constant, and this constraint is increasingly
felt as the number of base pair increases.

To compare the thermodynamics to a system where topological
constraints do not play a role we introduce a control system where the
hairpin loops have been opened by breaking the backbones between bases 10
and 11 of the first strand and 30 and 31 of the second strand, as shown in
Fig.~\ref{fig:schematics}(c). The free-energy profile of the control
shows the expected linear decrease in free energy for hybridization
in the absence of topological effects. It is also worth noting
that this system gains 30$k_{\rm B}T$ more in free energy from
hybridization than the hairpins do from forming the kissing complex. This
value is roughly the amount of free energy stored in the metastable kissing
complex (the most stable state for this system is the full duplex, since
the two sequences are fully complementary). In DNA nanotechnology systems
where such hairpins are used as fuels, this would be the amount of free
energy that is potentially available to do work.

It is also interesting to look at the structure of the kissing complex as
hybridization progresses. Fig.~\ref{fig:struct1} shows example structures
with different numbers of base pairs, and in Fig.~\ref{fig:dmms} the
probability that a given base is bound as a function of its position along
the loop is depicted for different numbers of total base pairs. For kissing
complexes with a few base pairs there is not a strong thermodynamic
preference for binding at a particular position in the loops, hence the
distribution in Fig.~\ref{fig:dmms} is roughly uniform. The exceptions are
the first and last bases in the loops for which base-pairing is
disfavoured, presumably because the more crowded environment that would
result makes binding at these positions entropically less favourable.

By contrast, for kissing complexes with the most favourable number of base
pairs ($N_{\rm tot}=14$), there is a very clear pattern for bonding. The
six bases closest to each stem are invariably base-paired, while the four
central bases have virtually no probability of binding. Therefore, the
structure of the ensemble of configurations with 14 base pairs are all very
similar to that in Fig.~\ref{fig:dmms}(b). We should also note that this
structure is very different from the typical schematics of kissing
complexes that tend to assume a single hybridized region, normally between
the central regions of the loops.

The reason for this well defined pattern of bonding becomes clear when we
examine the structure in more detail. It is simply the best way to maximise
the base pairing whilst satisfying the topological constraints. In
particular, the base pairing adjacent to each stem continues the two
helices formed by the stems (i.e.\ there is coaxial stacking) and there is
a parallel four-way junction\cite{Lilley00} at the coaxial stacking site
associated with the exchange of strands between the two helices.
Importantly, as the junction is parallel, the strand exchange leads to a
crossing with positive sign that helps to counterbalance the negative
crossings associated with each region of base pairing between the
loops~(Fig.~\ref{fig:struct1}(a)). This positive crossing comes with little
free energy cost because it does not lead to any loss of base pairing. By
contrast, the second positive crossing near to the centre of each hairpin
loop is associated with a reversal of the direction of wrapping of the two
chains around each other and is responsible for the inability of the two
loops to hybridize further without significant free energy cost.

The base pairing probability distributions for kissing complexes with 6 and
10 base pairs in Fig.~\ref{fig:dmms} illustrate how this tendency to base
pair at the extremes of the loops becomes more pronounced as the number of
base pairs is increased and the system becomes more topologically
constrained. However, the ensembles of such structures are still much more
diverse than for the fully formed kissing complex. Both the examples in
Fig.\ \ref{fig:struct1}(d) and (e) only have a single hybridized region,
with only the latter being adjacent to one of the hairpin stems.

The positive crossings of DNA strands associated with parallel four-way
junctions have previously been used to offset the negative crossings
associated with hybridization in so-called ``paranemic crossover''
motifs.\cite{Zhang02,Seeman01} In this motif hybridization occurs between
bubbles (a series of unpaired bases) in two duplexes leading to parallel
four way junctions at either end of the newly hybridized section (rather
than at just one end as for the hairpin loops). These junctions can exactly
offset half a turn in each of the helices that result, and so can lead to
complete hybridization between topologically closed species. These
paranemic crossover motifs have been proposed as an alternative to
``sticky'' single-stranded ends as a means for binding together different
molecules in DNA nanotechnology,~\cite{Zhang02,Shen04} and have been used
in making DNA triangles~\cite{Liu08b} and octahedra.~\cite{Shih04} Such
paranemic crossovers have also been shown to form between negatively
supercoiled homologous duplexes because their zero linking number helps to
alleviate the supercoiling.~\cite{Wang10}

Interestingly, a very similar structure to that depicted in
Fig.~\ref{fig:struct1}(b) has been identified for the inhibitory complex
between the antisense RNA CopA and its target messenger RNA
CopT~\cite{Kolb00,Kolb00b,Kolb01} and has also been suggested for other
such complexes.~\cite{Kolb01b} CopA and CopT have small hairpin loops that
associate to form an initial kissing complex, which then progresses to form
an ``extended'' kissing complex, where some of the base pairs in the
hairpin stems are lost in favour of two intermolecular helices that
coaxially stack with the rest of the hairpin stems at a parallel four-way
junction like that in Fig.~\ref{fig:struct1}(a). This progression is
dependent on the presence of bulges in the hairpin stems\cite{Kolb01} that
presumably aid the transformation by destabilizing the stems.
Intriguingly, the number of base pairs in the two intermolecular helices is
thought to be 15 with six bases in each of the loops connecting the ends of
these helices,\cite{Kolb00,Kolb00b} which is extremely similar to the
detailed structure of our kissing complexes.

Our results for the topologically unlinked system can be compared to the
experimental results of Refs.~\onlinecite{Boi05a} and~\onlinecite{Gre06a}
on the stability of these DNA kissing complexes.  Ref.~\onlinecite{Boi05a}
reported a high yeld (nearly 100\%) of kissing complexes at a single strand
concentration of $8\,\mu {\rm M}$ in a buffer of relatively high salt
concentration, while Ref.~\onlinecite{Gre06a} reported that kissing
complexes were not stable for their 21-base loop hairpins but at a
significantly lower strand concentration ($0.1\,\mu {\rm M}$) and at low
salt. The relative probability $\Phi$ of the system being in a bound state
(one or more interstrand base pairs) compared to an unbound state %(0
interstrand base pairs) can be inferred from the data in
Fig.~\ref{fig:fel2}. We find $\Phi \simeq 3100$.  Assuming high dilution,
this ratio can be extrapolated to a different simulation volume $v'$ simply
by dividing $\Phi$ by the ratio $v'/v$, where $v$ is the original
simulation volume. The relative probabilities of bound and unbound states
can than be related to the bulk yields $f_\infty$ as described in
Ref.~\onlinecite{Ouldridge_bulk_2010}. Extrapolating our results to the
conditions of Ref.~\onlinecite{Boi05a} we get $\Phi = 73.7$ and a bulk
yield $f_\infty=0.89$, which is consistent with the experimental result
that the kissing complexes were significantly more stable than the unbound
state.  By contrast, extrapolating our results to the concentration used in
Ref.~\onlinecite{Gre06a} we get $\Phi = 0.92$ and $f_\infty=0.37$,
indicating that the kissing complexes are less stable than the unbound
states, which is again consistent with the experimental findings,
especially when we take into account that our model is fitted to a much
higher salt concentration than that used in Ref.~\onlinecite{Gre06a} and
thus is expected to give an overestimate for the bulk yield in this case
since the electrostatic penalty for bringing two strands close
together would be larger at the experimental salt concentration.
It should also be pointed out that the sequences we study are
quite asymmetrical in GC content, but that our model does
not account for sequence-dependent effects. It is thus possible, depending on
temperature and salt concentration, that the resulting bonding pattern is
actually a single helix between the GC-rich regions of the loops.

\subsection{Topologically favoured complex; $Lk=-1$}
\begin{figure}[tb]
\begin{center}
\includegraphics[width=8.5cm]{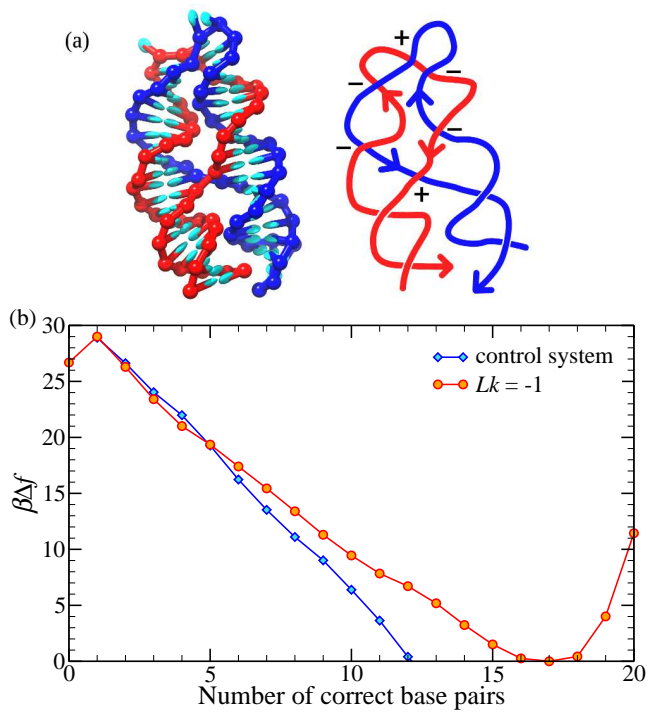} \\
\includegraphics[width=8.5cm]{fig5b}
\end{center}
\caption{\label{fig:fel1}
(a) Typical structure and topological sketch of the kissing complex with
$Lk=-1$ and (b) free energy profile associated with the formation with the
kissing complex, compared to that for the control system in
Fig.~\ref{fig:schematics}(c). To aid this comparison, the two free energy
profiles were set to have the same value at 1 base pair.
}
\end{figure}

We next consider topologically linked hairpins. Although they are less
experimentally relevant than the unlinked case, they nicely further
illustrate the effect of topology on hybridization. First we consider
hairpins with a linking number of -1 (Fig.~\ref{fig:schematics}(b)(ii)).
In this case the linkage has the same sense as the wrapping in duplex DNA
and we thus expect hybridization of the two hairpin loops to be easier than
for the unlinked case. The typical structure of the resulting kissing
complex and the free energy profile for hybridization are shown in
Fig.~\ref{fig:fel1}. There is a much lower entropic cost for initial
binding as compared to the topologically unlinked system, because the two
strands are already constrained to be close to each other due to the
linkage. The free energy profile also exhibits a much closer to linear
decrease with the number of base pairs formed than for the unlinked case,
and has a minimum at 17 base pairs.

The structure of the resulting kissing complex is quite similar to that for
the unlinked case in that it also has a parallel four-way junction at the
point where the two stems end. The reason for this structure is again that
the junction provides a positive crossing of the strands without any base
pairs being lost. The typical structure assumed by the kissing complex is
effectively two parallel helices and a small (2--4 base pairs) unbound
region where the strands bend back on themselves. The topological sketch in
Fig.\ \ref{fig:fel1}(a) shows how the two positive crossings (at the
four-way junction and at the end of the helices) and the linking number of
$-1$ allow the system to have four negative crossings, which is
topologically sufficient to form two double helical turns. Thus, that there
is still a free energy cost associated with the formation of the last base
pairs is due not so much to topological constraints
%--- the system has a sufficient number of positive crossings to allow full
%hybridisation --- 
but to geometric constraints arising because the backbones have to bend
around to bridge the two helices.

One interesting feature of the structure shown in Fig.~\ref{fig:fel1}(a) is
that the stems of both hairpins are only nine base pairs in length because
the hairpin loops have displaced one base pair from each stem. This then
raises the question of whether the four-way junction could migrate further
and lead to the opening of both hairpins. We note that there are a number
of features hindering the junction diffusion. Firstly, junction migration
is easiest when the junction adopts an ``open'' configuration where there
is no stacking across the junction,\cite{Panyutin94,Lilley00} rather than
the parallel stacked configuration typical of the kissing complex.
Secondly, the junction migration is resisted by the topology. If the total
number of base pairs is to remain constant during migration then the number
of base pairs in the duplex regions of the hybridized hairpin loops must
increase. However, as the linking number must also remain constant, this
also means that the unfavourable left-handed wrapping of the unhybridized
sections of the loops must increase. Our simulations corroborate this
picture. We observed that the position of the crossover between the end of
the two stems is rather stable, although it occasionally did move one or
two base pairs down.

\subsection{Topologically disfavoured complex; $Lk=+1$}
\begin{figure}[tb]
\begin{center}
\includegraphics[width=8.5cm]{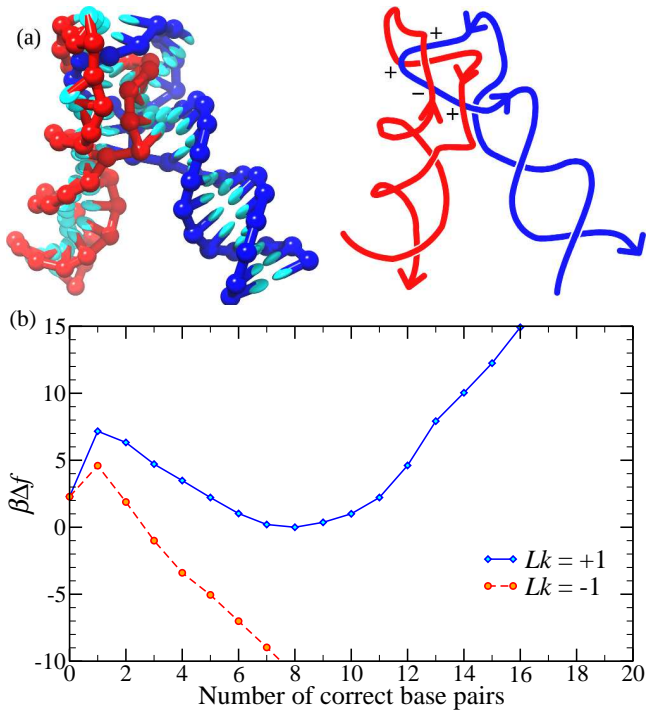}
\includegraphics[width=8.5cm]{fig6b}
\end{center}
\caption{\label{fig:fel0}
(a) Typical structure and topological sketch of the kissing complex with
$Lk=+1$ and (b) free energy profile associated with the formation of the
kissing complex, compared to that for $Lk=-1$.
%(Top) typical structure and topological sketch and (bottom) free energy
%profile of a topologically frustrated ($Lk=-1$) kissing complex. The most
%stable state is 8 hydrogen bonds between the two strands, approximately one
%half of the favourably linked case. The free energy cost of adding more
%hydrogen bonds is due to the difficulty of the two strands to wrap around
%in the direction opposite to that of the double helix to compensate the
%number of crossings.
}
\end{figure}

Finally, we consider topologically linked hairpins with a linking number of
+1 (Fig.~\ref{fig:schematics}(b)(iii)). In this case, the crossings
associated with the linkage are of the opposite sign to that for a duplex,
and so hinder base pairing between the loops. The effects of this
topological frustration are clear from the free energy profile in
Fig.~\ref{fig:fel0}. Now, the most stable kissing complex has only 8 base
pairs between the hairpin loops, and is only a few $k_{\rm B}T$ more stable
than the unhybridized state. Indeed, it is likely that for a slightly
shorter loops the topological frustration would be sufficient to totally
inhibit binding.

The effects of topology are underlined by the comparison with the
topologically favoured configuration with linking number $Lk=-1$, for which
a further 25$k_{\rm B}T$ drop in free energy is obtained on forming the
kissing complex. Visual inspection of the structure in Fig
\ref{fig:fel0}(a) indicates a much more distorted structure compared to the
previous cases. In this configuration, there is only a single negative
crossing (roughly enough for one half turn of the double helix), but three
positive crossings associated with the strands wrapping in the wrong sense
around each other.

\subsection{\label{sec:bb}Role of the backbone excluded volume}

Here, we consider how the results for kissing complexes depend on our
parametrization of the excluded volume interaction between backbone sites.
We do this firstly because this interaction term will play a key role in
determining how easy it is for the DNA chains to wrap around each other in
the wrong sense, and hence how many base pairs can be formed between
hairpin loops. But, secondly, in our original parametrization of the model
many of the properties to which we fitted are relatively insensitive to
this interaction, and so it is possible that this parameter does not have
its optimal value. For example, in the duplex state, in the high-salt
concentration to which our DNA model is fitted, the backbone sites are too
far away from each other for their mutual excluded volume to significantly
affect duplex properties. Other properties such as the single-stranded
persistence length played a greater role in its parametrization. The shape
of the interaction between backbone sites, modelled as a soft repulsion, is
also a significant approximation especially when two strands are close
together.

\begin{figure}[tb]
\begin{center}
\includegraphics[width=8.5cm]{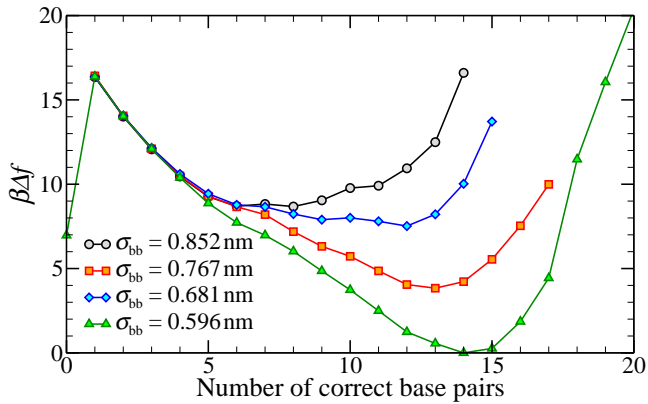}
\end{center}
\caption{\label{fig:bigs}
Effect of the backbone excluded volume on the free energy profiles for a
kissing complex with $Lk=0$. 
The original value of $\sigma_{\rm bb}$ is $0.596\,$nm.
}
\end{figure}

We have therefore repeated the calculation of the free energy profile for
the complex with $Lk=0$, but where we have changed the amount of repulsion
between backbones by increasing the effective radius $\sigma_{\rm bb}$ of
the coarse-grained backbone site by up to 30\%.
As shown in Fig.~\ref{fig:bigs}, as the repulsion is
increased, the average number of base pairs in the loop is diminished, and
the free energy gain for association is significantly lowered. Of course,
this change induces a large change the yield of kissing complexes at this
temperature.
Since our model's predictions for yields using the original value of
$\sigma_{\rm bb}$ are reasonably in line with the
experimental studies reported in Refs.~\onlinecite{Boi05a, Gre06a}, the
original parameterization appears to be robust.
Moreover, that the detailed pattern of base pairing is 
consistent with known structures
for RNA kissing complexes\cite{Bindewald08,Lee98a,Reb03a}
further corroborates this conclusion.

We also note that at the largest value studied for the range of the
repulsion the typical structure of the complex has a single intermolecular
helix. Since one could regard the increase in the range of the repulsion as
a very crude way to extrapolate the predictions of the model to a lower
salt concentration than the one at which it was parameterized, it is
possible that under those conditions binding between the hairpins' loops
involves a single intermolecular helix between the GC-rich regions.

\section{Conclusions}

Our simulations of the systems of kissing hairpins considered
experimentally by Bois {\it et al.}~\cite{Boi05a} using a recently
introduced coarse-grained DNA model clearly illustrate the effects of
topology (due to the constraint that the linking number must remain
constant) on the free energy landscape for the formation of a kissing
complex. For the unlinked case, this topological frustrations leads to
30\% of the bases being unpaired, and the binding free energy of the
kissing complex is significantly smaller than for a fully formed duplex
with equal strand length (it is equivalent to the binding of a duplex with
about 7 or 8 bases).

The free energy landscapes of the linked hairpins dramatically illustrate
how manipulation of the linking number can increase or decrease this
topological frustration. Compared to the unlinked system, the number of
base pairs in the most stable kissing complex increases by 3 in the
topologically more favourable case ($L_k=-1$) , but decreases by 6 in the
topologically less favourable case ($Lk=1$). Even though the two sequences
are fully complementary, the topological frustration also prevents the
stems being opened by a displacement reaction involving the propagation of
the intermolecular helices formed between the loops. It is this inhibition
of duplex hybridization that underlies the use of hairpins as fuel for
autonomous DNA nanodevices.
%[{\bf ADD something about catalyst: By contrast, a catalyst strand ... }]

The structure of the kissing complex is also of particular interest, as
this information is not straightforward to extract experimentally. We found
that the kissing complex has a somewhat unusual structure. In particular,
rather than having a single hybridized region between the loops, as might
have been anticipated, the kissing complex involves two intermolecular
helices that coaxially stack with the hairpin stems, and involves a
parallel four-way junction. This structure is favoured because there is a
positive crossing of the strands at the junction that helps to offset the
negative crossings associated with hybridization, but without any loss of
base pairing. By contrast, the positive crossing nearer the centre of the
loops is associated with unhybridized bases. For similar topological
reasons, parallel four-way junctions have also been observed for paranemic
motifs in which bulges in two separate duplexes cross hybridize.
Furthermore, a structure very similar to that reported here has also been
identified for an ``extended'' kissing complex between messenger and
antisense RNA that also involves parallel helices and a four-way
junction.~\cite{Kolb00,Kolb00b,Kolb01}

As with any study with a coarse-grained model, one needs to consider how
robust the results are and whether they might reflect any weaknesses of the
model. We explicitly checked this for the repulsion between backbone sites
in Section~\ref{sec:bb} and found that although the results could change
significantly when varying this parameter, our current value appears to be
physically most reasonable. In particular, the thermodynamics in our model
is consistent with the experimental stability of kissing complexes for
20-base hairpin loops in Refs.~\onlinecite{Boi05a,Gre06a} (after taking into
account differences in DNA and salt concentration). Furthermore, that
similar structures are seen for systems with similar topological
constraints suggests that our findings are physically robust.

 One of the approximations in the model that we should particularly
consider is the ``average base'' approximation, namely that that the bases
in our model have identical interaction properties, except that hydrogen
bonding can only occur between Watson-Crick base pairs. Although the G-C
content of the hairpin loops is close to half, 7 of those G-C base pairs
occur in one half of the loop. The consequences of this specific sequence
might be to make the kissing complex asymmetrical with a longer
intermolecular helix associated with the G-C rich half, or even to lead to
the total loss of the second more weakly bound helical section. In this
regard, it is interesting to note that the catalyst strand used by Bois
{\em et. al.} to open the kissing complex binds to that half of one of the
hairpins that is more weakly bound in the kissing complex~\cite{Boi05a}.

Although different topological states are only strictly well-defined for
closed-loop molecules, as we have shown here, topological effects can be
significant for linear DNA due to long-lived secondary structure that leads
to the formation of internal loops. These topological constraints can
inhibit hybridization and prevent the system reaching the lowest
free-energy state. DNA nanotechnology takes advantage of these effects when
using DNA hairpins as fuel for autonomous motors, but they could also
potentially be an obstacle to the successful self-assembly of DNA
nanostructures. For example if a strand hybridizes at its two ends to parts
of a second long strand, an internal loop results that will be potentially
restricted in its binding by topological effects, unless one of the already
hybridized ends unbinds, either due to melting or displacement. Therefore,
the longer the strands involved in a structure, the more likely that
topological constraints will have a significant effect on the ability of
the system to self assemble. This argument suggest that for DNA
origamis,~\cite{Rothemund06} the shortness of the staple strands (typically
having two or three binding domains) probably has the effect of reducing
the potential for topological effects to hinder self-assembly.
Furthermore, the excess of staple strands means that a
topologically-constrained bound strand can be displaced by one of the
equivalent staple strands from the reservoir in solution.

\section{Acknowledgments}
We would like to thank the Engineering and Physical Sciences Research Council 
for financial support.

%\bibliography{hairpin}

\begin{thebibliography}{10}%
\makeatletter
\providecommand \@ifxundefined [1]{%
 \ifx #1\undefined \expandafter \@firstoftwo
 \else \expandafter \@secondoftwo
\fi
}%
\providecommand \@ifnum [1]{%
 \ifnum #1\expandafter \@firstoftwo
 \else \expandafter \@secondoftwo
\fi
}%
\providecommand \enquote [1]{``#1''}%
\providecommand \bibnamefont  [1]{#1}%
\providecommand \bibfnamefont [1]{#1}%
\providecommand \citenamefont [1]{#1}%
\providecommand\href[0]{\@sanitize\@href}%
\providecommand\@href[1]{\endgroup\@@startlink{#1}\endgroup\@@href}%
\providecommand\@@href[1]{#1\@@endlink}%
\providecommand \@sanitize [0]{\begingroup\catcode`\&12\catcode`\#12\relax}%
\@ifxundefined \pdfoutput {\@firstoftwo}{%
 \@ifnum{\z@=\pdfoutput}{\@firstoftwo}{\@secondoftwo}%
}{%
 \providecommand\@@startlink[1]{\leavevmode}%
 \providecommand\@@endlink[0]{}%
}{%
 \providecommand\@@startlink[1]{%
  \leavevmode
  \pdfstartlink
   attr{/Border[0 0 1 ]/H/I/C[0 1 1]}%
   user{/Subtype/Link/A<</Type/Action/S/URI/URI(#1)>>}%
  \relax
 }%
 \providecommand\@@endlink[0]{\pdfendlink}%
}%
\providecommand \url  [0]{\begingroup\@sanitize \@url }%
\providecommand \@url [1]{\endgroup\@href {#1}{\urlprefix}}%
\providecommand \urlprefix [0]{URL }%
\providecommand \Eprint[0]{\href }%
\@ifxundefined \urlstyle {%
  \providecommand \doi [1]{doi:\discretionary{}{}{}#1}%
}{%
  \providecommand \doi [0]{doi:\discretionary{}{}{}\begingroup
  \urlstyle{rm}\Url }%
}%
\providecommand \doibase [0]{http://dx.doi.org/}%
\providecommand \Doi[1]{\href{\doibase#1}}%
\providecommand \selectlanguage [0]{\@gobble}%
\providecommand \bibinfo [0]{\@secondoftwo}%
\providecommand \bibfield [0]{\@secondoftwo}%
\providecommand \translation [1]{[#1]}%
\providecommand \BibitemOpen[0]{}%
\providecommand \bibitemStop [0]{}%
\providecommand \bibitemNoStop [0]{.\EOS\space}%
\providecommand \EOS [0]{\spacefactor3000\relax}%
\providecommand \BibitemShut [1]{\csname bibitem#1\endcsname}%
%</preamble>
\bibitem{Car09a}%
  \BibitemOpen
  \bibfield{author}{%
  \bibinfo {author} {\bibfnamefont{R.}~\bibnamefont{Carlson}},\ }%
  \bibfield{journal}{%
  \bibinfo {journal} {Nature Biotechnol.}\ }%
  \textbf{\bibinfo {volume} {27}},\ \bibinfo {pages} {1091} (\bibinfo {year}
  {2009})\BibitemShut{NoStop}%
\bibitem{See1982a}%
  \BibitemOpen
  \bibfield{author}{%
  \bibinfo {author} {\bibfnamefont{N.~C.}\ \bibnamefont{Seeman}},\ }%
  \bibfield{journal}{%
  \bibinfo {journal} {J. Theor. Bio.}\ }%
  \textbf{\bibinfo {volume} {99}},\ \bibinfo {pages} {237 } (\bibinfo {year}
  {1982})\BibitemShut{NoStop}%
\bibitem{Seeman2003}%
  \BibitemOpen
  \bibfield{author}{%
  \bibinfo {author} {\bibfnamefont{N.~C.}\ \bibnamefont{Seeman}},\ }%
  \bibfield{journal}{%
  \bibinfo {journal} {Nature}\ }%
  \textbf{\bibinfo {volume} {421}},\ \bibinfo {pages} {427} (\bibinfo {year}
  {2003})\BibitemShut{NoStop}%
\bibitem{Seeman10}%
  \BibitemOpen
  \bibfield{author}{%
  \bibinfo {author} {\bibfnamefont{N.~C.}\ \bibnamefont{Seeman}},\ }%
  \bibfield{journal}{%
  \bibinfo {journal} {Annu. Rev. Biochem.}\ }%
  \textbf{\bibinfo {volume} {79}},\ \bibinfo {pages} {65} (\bibinfo {year}
  {2010})\BibitemShut{NoStop}%
\bibitem{Bath2007}%
  \BibitemOpen
  \bibfield{author}{%
  \bibinfo {author} {\bibfnamefont{J.}~\bibnamefont{Bath}}\ and\ \bibinfo
  {author} {\bibfnamefont{A.~J.}\ \bibnamefont{Turberfield}},\ }%
  \bibfield{journal}{%
  \bibinfo {journal} {Nat. Nanotechnol.}\ }%
  \textbf{\bibinfo {volume} {2}},\ \bibinfo {pages} {275} (\bibinfo {year}
  {2007})\BibitemShut{NoStop}%
\bibitem{Varani95}%
  \BibitemOpen
  \bibfield{author}{%
  \bibinfo {author} {\bibfnamefont{G.}~\bibnamefont{Varani}},\ }%
  \bibfield{journal}{%
  \bibinfo {journal} {Annu. Rev. Biophys. Biomol. Struct.}\ }%
  \textbf{\bibinfo {volume} {24}},\ \bibinfo {pages} {379} (\bibinfo {year}
  {1995})\BibitemShut{NoStop}%
\bibitem{Bikard10}%
  \BibitemOpen
  \bibfield{author}{%
  \bibinfo {author} {\bibfnamefont{D.}~\bibnamefont{Bikard}}, \bibinfo {author}
  {\bibfnamefont{C.}~\bibnamefont{Loot}}, \bibinfo {author}
  {\bibfnamefont{Z.}~\bibnamefont{Baharoglu}},\ and\ \bibinfo {author}
  {\bibfnamefont{D.}~\bibnamefont{Mazel}},\ }%
  \bibfield{journal}{%
  \bibinfo {journal} {Microbiol. Mol. Biol. Rev.}\ }%
  \textbf{\bibinfo {volume} {74}},\ \bibinfo {pages} {570} (\bibinfo {year}
  {2010})\BibitemShut{NoStop}%
\bibitem{GlucksmannKuis96}%
  \BibitemOpen
  \bibfield{author}{%
  \bibinfo {author} {\bibfnamefont{M.~A.}\ \bibnamefont{Glucksmann-Kuis}},
  \bibinfo {author} {\bibfnamefont{X.}~\bibnamefont{Dai}}, \bibinfo {author}
  {\bibfnamefont{P.}~\bibnamefont{Markiewicz}},\ and\ \bibinfo {author}
  {\bibfnamefont{L.}~\bibnamefont{Rothman-Denes}},\ }%
  \bibfield{journal}{%
  \bibinfo {journal} {Cell}\ }%
  \textbf{\bibinfo {volume} {84}},\ \bibinfo {pages} {147} (\bibinfo {year}
  {1996})\BibitemShut{NoStop}%
\bibitem{Oettinger04}%
  \BibitemOpen
  \bibfield{author}{%
  \bibinfo {author} {\bibfnamefont{M.~A.}\ \bibnamefont{Oettinger}},\ }%
  \bibfield{journal}{%
  \bibinfo {journal} {Nature}\ }%
  \textbf{\bibinfo {volume} {432}},\ \bibinfo {pages} {960} (\bibinfo {year}
  {2004})\BibitemShut{NoStop}%
\bibitem{Santhana96}%
  \BibitemOpen
  \bibfield{author}{%
  \bibinfo {author} {\bibfnamefont{S.~V.}\ \bibnamefont{Santhana~Mariappan}},
  \bibinfo {author} {\bibfnamefont{A.~E.}\ \bibnamefont{Garcia}},\ and\
  \bibinfo {author} {\bibfnamefont{G.}~\bibnamefont{Gupta}},\ }%
  \bibfield{journal}{%
  \bibinfo {journal} {Nucl. Acids Res.}\ }%
  \textbf{\bibinfo {volume} {24}},\ \bibinfo {pages} {775} (\bibinfo {year}
  {1996})\BibitemShut{NoStop}%
\bibitem{Lam11}%
  \BibitemOpen
  \bibfield{author}{%
  \bibinfo {author} {\bibfnamefont{S.~L.}\ \bibnamefont{Lam}}, \bibinfo
  {author} {\bibfnamefont{F.}~\bibnamefont{Wu}}, \bibinfo {author}
  {\bibfnamefont{H.}~\bibnamefont{Yang}},\ and\ \bibinfo {author}
  {\bibfnamefont{L.~M.}\ \bibnamefont{Chi}},\ }%
  \bibfield{journal}{%
  \bibinfo {journal} {Nucl. Acids Res.}\ }%
  \textbf{\bibinfo {volume} {39}},\ \bibinfo {pages} {6260} (\bibinfo {year}
  {2011})\BibitemShut{NoStop}%
\bibitem{Dirks04}%
  \BibitemOpen
  \bibfield{author}{%
  \bibinfo {author} {\bibfnamefont{R.~M.}\ \bibnamefont{Dirks}}\ and\ \bibinfo
  {author} {\bibfnamefont{N.~A.}\ \bibnamefont{Pierce}},\ }%
  \bibfield{journal}{%
  \bibinfo {journal} {Proc. Natl. Acad. Sci. USA}\ }%
  \textbf{\bibinfo {volume} {101}},\ \bibinfo {pages} {15275} (\bibinfo {year}
  {2004})\BibitemShut{NoStop}%
\bibitem{Boi05a}%
  \BibitemOpen
  \bibfield{author}{%
  \bibinfo {author} {\bibfnamefont{J.~S.}\ \bibnamefont{Bois}}, \bibinfo
  {author} {\bibfnamefont{S.}~\bibnamefont{Venkataraman}}, \bibinfo {author}
  {\bibfnamefont{H.~M.~T.}\ \bibnamefont{Choi}}, \bibinfo {author}
  {\bibfnamefont{A.~J.}\ \bibnamefont{Spakowitz}}, \bibinfo {author}
  {\bibfnamefont{Z.-G.}\ \bibnamefont{Wang}},\ and\ \bibinfo {author}
  {\bibfnamefont{N.~A.}\ \bibnamefont{Pierce}},\ }%
  \bibfield{journal}{%
  \bibinfo {journal} {Nucl. Acids Res.}\ }%
  \textbf{\bibinfo {volume} {33}},\ \bibinfo {pages} {4090} (\bibinfo {year}
  {2005})\BibitemShut{NoStop}%
\bibitem{Gre06a}%
  \BibitemOpen
  \bibfield{author}{%
  \bibinfo {author} {\bibfnamefont{S.~J.}\ \bibnamefont{Green}}, \bibinfo
  {author} {\bibfnamefont{D.}~\bibnamefont{Lubrich}},\ and\ \bibinfo {author}
  {\bibfnamefont{A.~J.}\ \bibnamefont{Turberfield}},\ }%
  \bibfield{journal}{%
  \bibinfo {journal} {Biophys. J.}\ }%
  \textbf{\bibinfo {volume} {91}},\ \bibinfo {pages} {2966 } (\bibinfo {year}
  {2006})\BibitemShut{NoStop}%
\bibitem{Venkataraman2007}%
  \BibitemOpen
  \bibfield{author}{%
  \bibinfo {author} {\bibfnamefont{S.}~\bibnamefont{Venkataraman}}, \bibinfo
  {author} {\bibfnamefont{R.~M.}\ \bibnamefont{Dirks}}, \bibinfo {author}
  {\bibfnamefont{P.~W.~K.}\ \bibnamefont{Rothemund}}, \bibinfo {author}
  {\bibfnamefont{E.}~\bibnamefont{Winfree}},\ and\ \bibinfo {author}
  {\bibfnamefont{N.~A.}\ \bibnamefont{Pierce}},\ }%
  \bibfield{journal}{%
  \bibinfo {journal} {Nat. Nanotechnol.}\ }%
  \textbf{\bibinfo {volume} {2}},\ \bibinfo {pages} {490} (\bibinfo {year}
  {2007})\BibitemShut{NoStop}%
\bibitem{Green2008}%
  \BibitemOpen
  \bibfield{author}{%
  \bibinfo {author} {\bibfnamefont{S.~J.}\ \bibnamefont{Green}}, \bibinfo
  {author} {\bibfnamefont{J.}~\bibnamefont{Bath}},\ and\ \bibinfo {author}
  {\bibfnamefont{A.~J.}\ \bibnamefont{Turberfield}},\ }%
  \bibfield{journal}{%
  \bibinfo {journal} {Phys. Rev. Lett.}\ }%
  \textbf{\bibinfo {volume} {101}},\ \bibinfo {pages} {238101} (\bibinfo {year}
  {2008})\BibitemShut{NoStop}%
\bibitem{Yin2008}%
  \BibitemOpen
  \bibfield{author}{%
  \bibinfo {author} {\bibfnamefont{P.}~\bibnamefont{Yin}}, \bibinfo {author}
  {\bibfnamefont{H.~M.}\ \bibnamefont{Choi}}, \bibinfo {author}
  {\bibfnamefont{C.~R.}\ \bibnamefont{Calvert}},\ and\ \bibinfo {author}
  {\bibfnamefont{N.~A.}\ \bibnamefont{Pierce}},\ }%
  \bibfield{journal}{%
  \bibinfo {journal} {Nature}\ }%
  \textbf{\bibinfo {volume} {451}},\ \bibinfo {pages} {318} (\bibinfo {year}
  {2008})\BibitemShut{NoStop}%
\bibitem{Muscat2011}%
  \BibitemOpen
  \bibfield{author}{%
  \bibinfo {author} {\bibfnamefont{R.~A.}\ \bibnamefont{Muscat}}, \bibinfo
  {author} {\bibfnamefont{J.}~\bibnamefont{Bath}},\ and\ \bibinfo {author}
  {\bibfnamefont{A.~J.}\ \bibnamefont{Turberfield}},\ }%
  \bibfield{journal}{%
  \bibinfo {journal} {Nano Lett.}\ }%
  \textbf{\bibinfo {volume} {11}},\ \bibinfo {pages} {982} (\bibinfo {year}
  {2011})\BibitemShut{NoStop}%
\bibitem{Ouldridge_tweezers_2010}%
  \BibitemOpen
  \bibfield{author}{%
  \bibinfo {author} {\bibfnamefont{T.~E.}\ \bibnamefont{Ouldridge}}, \bibinfo
  {author} {\bibfnamefont{A.~A.}\ \bibnamefont{Louis}},\ and\ \bibinfo {author}
  {\bibfnamefont{J.~P.~K.}\ \bibnamefont{Doye}},\ }%
  \bibfield{journal}{%
  \bibinfo {journal} {Phys. Rev. Lett.}\ }%
  \textbf{\bibinfo {volume} {104}},\ \bibinfo {pages} {178101} (\bibinfo {year}
  {2010})\BibitemShut{NoStop}%
\bibitem{Tin99a}%
  \BibitemOpen
  \bibfield{author}{%
  \bibinfo {author} {\bibfnamefont{I.}~\bibnamefont{Tinoco~Jr}}\ and\ \bibinfo
  {author} {\bibfnamefont{C.}~\bibnamefont{Bustamante}},\ }%
  \bibfield{journal}{%
  \bibinfo {journal} {J. Mol. Biol.}\ }%
  \textbf{\bibinfo {volume} {293}},\ \bibinfo {pages} {271 } (\bibinfo {year}
  {1999})\BibitemShut{NoStop}%
\bibitem{Bois9}%
  \BibitemOpen
  \bibfield{author}{%
  \bibinfo {author} {\bibfnamefont{C.}~\bibnamefont{Brunel}}, \bibinfo {author}
  {\bibfnamefont{R.}~\bibnamefont{Marquet}}, \bibinfo {author}
  {\bibfnamefont{P.}~\bibnamefont{Romby}},\ and\ \bibinfo {author}
  {\bibfnamefont{C.}~\bibnamefont{Ehresmann}},\ }%
  \bibfield{journal}{%
  \bibinfo {journal} {Biochimie}\ }%
  \textbf{\bibinfo {volume} {84}},\ \bibinfo {pages} {925} (\bibinfo {year}
  {2002})\BibitemShut{NoStop}%
\bibitem{Bindewald08}%
  \BibitemOpen
  \bibfield{author}{%
  \bibinfo {author} {\bibfnamefont{E.}~\bibnamefont{Bindewald}}, \bibinfo
  {author} {\bibfnamefont{R.}~\bibnamefont{Hayes}}, \bibinfo {author}
  {\bibfnamefont{Y.~G.}\ \bibnamefont{Yingling}}, \bibinfo {author}
  {\bibfnamefont{W.}~\bibnamefont{Kasprzak}},\ and\ \bibinfo {author}
  {\bibfnamefont{B.~A.}\ \bibnamefont{Shapiro}},\ }%
  \bibfield{journal}{%
  \bibinfo {journal} {Nucl. Acids. Res.}\ }%
  \textbf{\bibinfo {volume} {36}},\ \bibinfo {pages} {D392} (\bibinfo {year}
  {2008})\BibitemShut{NoStop}%
\bibitem{Lee98a}%
  \BibitemOpen
  \bibfield{author}{%
  \bibinfo {author} {\bibfnamefont{A.~J.}\ \bibnamefont{Lee}}\ and\ \bibinfo
  {author} {\bibfnamefont{D.~M.}\ \bibnamefont{Crothers}},\ }%
  \bibfield{journal}{%
  \bibinfo {journal} {Structure}\ }%
  \textbf{\bibinfo {volume} {6}},\ \bibinfo {pages} {993 } (\bibinfo {year}
  {1998})\BibitemShut{NoStop}%
\bibitem{Reb03a}%
  \BibitemOpen
  \bibfield{author}{%
  \bibinfo {author} {\bibfnamefont{K.}~\bibnamefont{R{\'e}blov{\'a}}}, \bibinfo
  {author} {\bibfnamefont{N.}~\bibnamefont{{\v S}pa{\v c}kov{\' a}}}, \bibinfo
  {author} {\bibfnamefont{J.~E.}\ \bibnamefont{{\v S}poner}}, \bibinfo {author}
  {\bibfnamefont{J.}~\bibnamefont{Ko{\v c}a}},\ and\ \bibinfo {author}
  {\bibfnamefont{J.}~\bibnamefont{{\v S}poner}},\ }%
  \bibfield{journal}{%
  \bibinfo {journal} {Nucl. Acids Res.}\ }%
  \textbf{\bibinfo {volume} {31}},\ \bibinfo {pages} {6942} (\bibinfo {year}
  {2003})\BibitemShut{NoStop}%
\bibitem{Li06}%
  \BibitemOpen
  \bibfield{author}{%
  \bibinfo {author} {\bibfnamefont{P.~T.~X.}\ \bibnamefont{Li}}, \bibinfo
  {author} {\bibfnamefont{C.}~\bibnamefont{Bustamente}},\ and\ \bibinfo
  {author} {\bibfnamefont{I.}~\bibnamefont{Tinoco~Jr.}},\ }%
  \bibfield{journal}{%
  \bibinfo {journal} {Proc. Natl. Acad. Sci. USA}\ }%
  \textbf{\bibinfo {volume} {103}},\ \bibinfo {pages} {15847} (\bibinfo {year}
  {2006})\BibitemShut{NoStop}%
\bibitem{Li09d}%
  \BibitemOpen
  \bibfield{author}{%
  \bibinfo {author} {\bibfnamefont{P.~T.~X.}\ \bibnamefont{Li}}\ and\ \bibinfo
  {author} {\bibfnamefont{I.}~\bibnamefont{Tinoco~Jr.}},\ }%
  \bibfield{journal}{%
  \bibinfo {journal} {J. Mol. Biol.}\ }%
  \textbf{\bibinfo {volume} {386}},\ \bibinfo {pages} {1343} (\bibinfo {year}
  {2009})\BibitemShut{NoStop}%
\bibitem{Horiya03}%
  \BibitemOpen
  \bibfield{author}{%
  \bibinfo {author} {\bibfnamefont{S.}~\bibnamefont{Horiya}}, \bibinfo {author}
  {\bibfnamefont{X.}~\bibnamefont{Li}}, \bibinfo {author}
  {\bibfnamefont{G.}~\bibnamefont{Kawai}}, \bibinfo {author}
  {\bibfnamefont{R.}~\bibnamefont{Saito}}, \bibinfo {author}
  {\bibfnamefont{A.}~\bibnamefont{Katoh}}, \bibinfo {author}
  {\bibfnamefont{K.}~\bibnamefont{Kobayashi}},\ and\ \bibinfo {author}
  {\bibfnamefont{K.}~\bibnamefont{Harada}},\ }%
  \bibfield{journal}{%
  \bibinfo {journal} {Chem. Biol.}\ }%
  \textbf{\bibinfo {volume} {10}},\ \bibinfo {pages} {645} (\bibinfo {year}
  {2003})\BibitemShut{NoStop}%
\bibitem{Severcan10}%
  \BibitemOpen
  \bibfield{author}{%
  \bibinfo {author} {\bibfnamefont{I.}~\bibnamefont{Severcan}}, \bibinfo
  {author} {\bibfnamefont{C.}~\bibnamefont{Geary}}, \bibinfo {author}
  {\bibfnamefont{A.}~\bibnamefont{Chworos}}, \bibinfo {author}
  {\bibfnamefont{N.}~\bibnamefont{Voss}}, \bibinfo {author}
  {\bibfnamefont{E.}~\bibnamefont{Jacovetty}},\ and\ \bibinfo {author}
  {\bibfnamefont{L.}~\bibnamefont{Jaeger}},\ }%
  \bibfield{journal}{%
  \bibinfo {journal} {Nature Chem.}\ }%
  \textbf{\bibinfo {volume} {2}},\ \bibinfo {pages} {772} (\bibinfo {year}
  {2010})\BibitemShut{NoStop}%
\bibitem{Grabow11}%
  \BibitemOpen
  \bibfield{author}{%
  \bibinfo {author} {\bibfnamefont{W.~W.}\ \bibnamefont{Grabow}}, \bibinfo
  {author} {\bibfnamefont{P.}~\bibnamefont{Zakrevsky}}, \bibinfo {author}
  {\bibfnamefont{K.~A.}\ \bibnamefont{Afonin}}, \bibinfo {author}
  {\bibfnamefont{A.}~\bibnamefont{Chworos}}, \bibinfo {author}
  {\bibfnamefont{B.~A.}\ \bibnamefont{Shapiro}},\ and\ \bibinfo {author}
  {\bibfnamefont{L.}~\bibnamefont{Jaeger}},\ }%
  \bibfield{journal}{%
  \bibinfo {journal} {Nano. Lett.}\ }%
  \textbf{\bibinfo {volume} {11}},\ \bibinfo {pages} {878} (\bibinfo {year}
  {2011})\BibitemShut{NoStop}%
\bibitem{Barbault02}%
  \BibitemOpen
  \bibfield{author}{%
  \bibinfo {author} {\bibfnamefont{F.}~\bibnamefont{Barbault}}, \bibinfo
  {author} {\bibfnamefont{T.}~\bibnamefont{Huynh-Dinh}}, \bibinfo {author}
  {\bibfnamefont{J.}~\bibnamefont{Paoletti}},\ and\ \bibinfo {author}
  {\bibfnamefont{G.}~\bibnamefont{Lancelot}},\ }%
  \bibfield{journal}{%
  \bibinfo {journal} {J. Biomol. Struct. Dyn.}\ }%
  \textbf{\bibinfo {volume} {19}},\ \bibinfo {pages} {649} (\bibinfo {year}
  {2002})\BibitemShut{NoStop}%
\bibitem{Turberfield2003}%
  \BibitemOpen
  \bibfield{author}{%
  \bibinfo {author} {\bibfnamefont{A.~J.}\ \bibnamefont{Turberfield}}, \bibinfo
  {author} {\bibfnamefont{J.~C.}\ \bibnamefont{Mitchell}}, \bibinfo {author}
  {\bibfnamefont{B.}~\bibnamefont{Yurke}}, \bibinfo {author}
  {\bibfnamefont{A.~P.}\ \bibnamefont{Mills}}, \bibinfo {author}
  {\bibfnamefont{M.~I.}\ \bibnamefont{Blakey}},\ and\ \bibinfo {author}
  {\bibfnamefont{F.~C.}\ \bibnamefont{Simmel}},\ }%
  \bibfield{journal}{%
  \bibinfo {journal} {Phys. Rev. Lett.}\ }%
  \textbf{\bibinfo {volume} {90}},\ \bibinfo {pages} {118102} (\bibinfo {year}
  {2003})\BibitemShut{NoStop}%
\bibitem{See06a}%
  \BibitemOpen
  \bibfield{author}{%
  \bibinfo {author} {\bibfnamefont{G.}~\bibnamefont{Seelig}}, \bibinfo {author}
  {\bibfnamefont{B.}~\bibnamefont{Yurke}},\ and\ \bibinfo {author}
  {\bibfnamefont{E.}~\bibnamefont{Winfree}},\ }%
  \bibfield{journal}{%
  \bibinfo {journal} {J. Am. Chem. Soc.}\ }%
  \textbf{\bibinfo {volume} {128}},\ \bibinfo {pages} {12211} (\bibinfo {year}
  {2006})\BibitemShut{NoStop}%
\bibitem{Yurke2000}%
  \BibitemOpen
  \bibfield{author}{%
  \bibinfo {author} {\bibfnamefont{B.}~\bibnamefont{Yurke}}, \bibinfo {author}
  {\bibfnamefont{A.~J.}\ \bibnamefont{Turberfield}}, \bibinfo {author}
  {\bibfnamefont{A.~P.}\ \bibnamefont{Mills}}, \bibinfo {author}
  {\bibfnamefont{F.~C.}\ \bibnamefont{Simmel}},\ and\ \bibinfo {author}
  {\bibfnamefont{J.}~\bibnamefont{Neumann}},\ }%
  \bibfield{journal}{%
  \bibinfo {journal} {Nature}\ }%
  \textbf{\bibinfo {volume} {406}},\ \bibinfo {pages} {605} (\bibinfo {year}
  {2000})\BibitemShut{NoStop}%
\bibitem{Ouldridge_long_2011}%
  \BibitemOpen
  \bibfield{author}{%
  \bibinfo {author} {\bibfnamefont{T.~E.}\ \bibnamefont{Ouldridge}}, \bibinfo
  {author} {\bibfnamefont{A.~A.}\ \bibnamefont{Louis}},\ and\ \bibinfo {author}
  {\bibfnamefont{J.~P.~K.}\ \bibnamefont{Doye}},\ }%
  \bibfield{journal}{%
  \bibinfo {journal} {J. Chem. Phys.}\ }%
  \textbf{\bibinfo {volume} {134}},\ \bibinfo {pages} {085101} (\bibinfo {year}
  {2011})\BibitemShut{NoStop}%
\bibitem{Ouldridge11b}%
  \BibitemOpen
  \bibfield{author}{%
  \bibinfo {author} {\bibfnamefont{T.~E.}\ \bibnamefont{Ouldridge}},\ }%
  \emph{\bibinfo {title} {Coarse-grained modelling of DNA and DNA
  self-assembly}},\ Ph.D. thesis,\ \bibinfo {school} {University of Oxford}
  (\bibinfo {year} {2011})\BibitemShut{NoStop}%
\bibitem{Ouldridge12}%
  \BibitemOpen
  \bibfield{author}{%
  \bibinfo {author} {\bibfnamefont{T.~E.}\ \bibnamefont{Ouldridge}}, \bibinfo
  {author} {\bibfnamefont{A.~A.}\ \bibnamefont{Louis}},\ and\ \bibinfo {author}
  {\bibfnamefont{J.~P.~K.}\ \bibnamefont{Doye}},\ }%
  \bibinfo {note} {in preparation}\BibitemShut{NoStop}%
\bibitem{CDM}%
  \BibitemOpen
  \bibfield{author}{%
  \bibinfo {author} {\bibfnamefont{C.}~\bibnamefont{{De Michele}}}, \bibinfo
  {author} {\bibfnamefont{L.}~\bibnamefont{Rovigatti}}, \bibinfo{author} {\bibfnamefont{T.}~\bibnamefont{Bellini}}\ and\ \bibinfo {author}
  {\bibfnamefont{F.}~\bibnamefont{Sciortino}},\ }%
  \bibfield{journal}{%
  \bibinfo {journal} {in preparation}}%
   (\bibinfo {year} {2012})\BibitemShut{NoStop}%
\bibitem{Bath2005}%
  \BibitemOpen
  \bibfield{author}{%
  \bibinfo {author} {\bibfnamefont{J.}~\bibnamefont{Bath}}, \bibinfo {author}
  {\bibfnamefont{S.~J.}\ \bibnamefont{Green}},\ and\ \bibinfo {author}
  {\bibfnamefont{A.~J.}\ \bibnamefont{Turberfield}},\ }%
  \bibfield{journal}{%
  \bibinfo {journal} {Angew. Chem. Int. Ed.}\ }%
  \textbf{\bibinfo {volume} {117}},\ \bibinfo {pages} {4432} (\bibinfo {year}
  {2005})\BibitemShut{NoStop}%
\bibitem{Bath2009}%
  \BibitemOpen
  \bibfield{author}{%
  \bibinfo {author} {\bibfnamefont{J.}~\bibnamefont{Bath}}, \bibinfo {author}
  {\bibfnamefont{S.~J.}\ \bibnamefont{Green}}, \bibinfo {author}
  {\bibfnamefont{K.~E.}\ \bibnamefont{Allan}},\ and\ \bibinfo {author}
  {\bibfnamefont{A.~J.}\ \bibnamefont{Turberfield}},\ }%
  \bibfield{journal}{%
  \bibinfo {journal} {Small}\ }%
  \textbf{\bibinfo {volume} {5}},\ \bibinfo {pages} {1513} (\bibinfo {year}
  {2009})\BibitemShut{NoStop}%
\bibitem{Zhang_disp_2009}%
  \BibitemOpen
  \bibfield{author}{%
  \bibinfo {author} {\bibfnamefont{D.}~\bibnamefont{Zhang}}\ and\ \bibinfo
  {author} {\bibfnamefont{E.}~\bibnamefont{Winfree}},\ }%
  \bibfield{journal}{%
  \bibinfo {journal} {J. Am. Chem. Soc.}\ }%
  \textbf{\bibinfo {volume} {131}},\ \bibinfo {pages} {17303} (\bibinfo {year}
  {2009})\BibitemShut{NoStop}%
\bibitem{Smith96}%
  \BibitemOpen
  \bibfield{author}{%
  \bibinfo {author} {\bibfnamefont{S.~B.}\ \bibnamefont{Smith}}, \bibinfo
  {author} {\bibfnamefont{Y.}~\bibnamefont{Cui}},\ and\ \bibinfo {author}
  {\bibfnamefont{C.}~\bibnamefont{Bustamente}},\ }%
  \bibfield{journal}{%
  \bibinfo {journal} {Nature}\ }%
  \textbf{\bibinfo {volume} {271}},\ \bibinfo {pages} {795} (\bibinfo {year}
  {1996})\BibitemShut{NoStop}%
\bibitem{Ramreddy11}%
  \BibitemOpen
  \bibfield{author}{%
  \bibinfo {author} {\bibfnamefont{T.}~\bibnamefont{Ramredday}}, \bibinfo
  {author} {\bibfnamefont{R.}~\bibnamefont{Sachidanandam}},\ and\ \bibinfo
  {author} {\bibfnamefont{T.~R.}\ \bibnamefont{Strick}},\ }%
  \bibfield{journal}{%
  \bibinfo {journal} {Nucl. Acids Res.}\ }%
  \textbf{\bibinfo {volume} {39}},\ \bibinfo {pages} {4275} (\bibinfo {year}
  {2011})\BibitemShut{NoStop}%
\bibitem{BelliniScience}%
  \BibitemOpen
  \bibfield{author}{%
  \bibinfo {author} {\bibfnamefont{M.}~\bibnamefont{Nakata}}, \bibinfo {author}
  {\bibfnamefont{G.}~\bibnamefont{Zanchetta}}, \bibinfo {author}
  {\bibfnamefont{B.~D.}\ \bibnamefont{Chapman}}, \bibinfo {author}
  {\bibfnamefont{C.~D.}\ \bibnamefont{Jones}}, \bibinfo {author}
  {\bibfnamefont{J.~O.}\ \bibnamefont{Cross}}, \bibinfo {author}
  {\bibfnamefont{R.}~\bibnamefont{Pindak}}, \bibinfo {author}
  {\bibfnamefont{T.}~\bibnamefont{Bellini}},\ and\ \bibinfo {author}
  {\bibfnamefont{N.~A.}\ \bibnamefont{Clark}},\ }%
  \bibfield{journal}{%
  \Doi{10.1126/science.1143826}{\bibinfo {journal} {Science}}\ }%
  \textbf{\bibinfo {volume} {318}},\ \bibinfo {pages} {1276} (\bibinfo {year}
  {2007})\BibitemShut{NoStop}%
\bibitem{Whitelam2007}%
  \BibitemOpen
  \bibfield{author}{%
  \bibinfo {author} {\bibfnamefont{S.}~\bibnamefont{Whitelam}}\ and\ \bibinfo
  {author} {\bibfnamefont{P.~L.}\ \bibnamefont{Geissler}},\ }%
  \bibfield{journal}{%
  \bibinfo {journal} {J. Chem. Phys.}\ }%
  \textbf{\bibinfo {volume} {127}},\ \bibinfo {pages} {154101} (\bibinfo {year}
  {2007})\BibitemShut{NoStop}%
\bibitem{Whitelam2009}%
  \BibitemOpen
  \bibfield{author}{%
  \bibinfo {author} {\bibfnamefont{S.}~\bibnamefont{Whitelam}}, \bibinfo
  {author} {\bibfnamefont{E.~H.}\ \bibnamefont{Feng}}, \bibinfo {author}
  {\bibfnamefont{M.~F.}\ \bibnamefont{Hagan}},\ and\ \bibinfo {author}
  {\bibfnamefont{P.~L.}\ \bibnamefont{Geissler}},\ }%
  \bibfield{journal}{%
  \bibinfo {journal} {Soft Matter}\ }%
  \textbf{\bibinfo {volume} {5}},\ \bibinfo {pages} {1521} (\bibinfo {year}
  {2009})\BibitemShut{NoStop}%
\bibitem{Torrie1977}%
  \BibitemOpen
  \bibfield{author}{%
  \bibinfo {author} {\bibfnamefont{G.~M.}\ \bibnamefont{Torrie}}\ and\ \bibinfo
  {author} {\bibfnamefont{J.~P.}\ \bibnamefont{Valleau}},\ }%
  \bibfield{journal}{%
  \bibinfo {journal} {J. Comp. Phys.}\ }%
  \textbf{\bibinfo {volume} {23}},\ \bibinfo {pages} {187} (\bibinfo {year}
  {1977})\BibitemShut{NoStop}%
\bibitem{Kumar1992}%
  \BibitemOpen
  \bibfield{author}{%
  \bibinfo {author} {\bibfnamefont{S.}~\bibnamefont{Kumar}}, \bibinfo {author}
  {\bibfnamefont{J.~M.}\ \bibnamefont{Rosenberg}}, \bibinfo {author}
  {\bibfnamefont{D.}~\bibnamefont{Bouzida}}, \bibinfo {author}
  {\bibfnamefont{R.~H.}\ \bibnamefont{Swendsen}},\ and\ \bibinfo {author}
  {\bibfnamefont{P.~A.}\ \bibnamefont{Kollman}},\ }%
  \bibfield{journal}{%
  \bibinfo {journal} {J. Comput. Chem.}\ }%
  \textbf{\bibinfo {volume} {13}},\ \bibinfo {pages} {1011} (\bibinfo {year}
  {1992})\BibitemShut{NoStop}%
\bibitem{DNAtopology_signs}%
  \BibitemOpen
  \bibinfo {note} {Sometimes an alternative convention is used where the
  strands in a DNA duplex are considered to be parallel, so that positive
  linking numbers result (Ref.~\onlinecite{DNAtopology})}\BibitemShut{NoStop}%
\bibitem{DNAtopology}%
  \BibitemOpen
  \bibfield{author}{%
  \bibinfo {author} {\bibfnamefont{A.~D.}\ \bibnamefont{Bates}}\ and\ \bibinfo
  {author} {\bibfnamefont{A.}~\bibnamefont{Maxwell}},\ }%
  \emph{\bibinfo {title} {DNA Topology}}\ (\bibinfo {publisher} {Oxford
  University Press},\ \bibinfo {year} {2005})
  \BibitemShut{NoStop}%
\bibitem{Lilley00}%
  \BibitemOpen
  \bibfield{author}{%
  \bibinfo {author} {\bibfnamefont{D.~M.~J.}\ \bibnamefont{Lilley}},\ }%
  \bibfield{journal}{%
  \bibinfo {journal} {Quart. Rev. Biophys.}\ }%
  \textbf{\bibinfo {volume} {33}},\ \bibinfo {pages} {109} (\bibinfo {year}
  {2000})\BibitemShut{NoStop}%
\bibitem{Zhang02}%
  \BibitemOpen
  \bibfield{author}{%
  \bibinfo {author} {\bibfnamefont{X.}~\bibnamefont{Zhang}}, \bibinfo {author}
  {\bibfnamefont{H.}~\bibnamefont{Yan}}, \bibinfo {author}
  {\bibfnamefont{Z.}~\bibnamefont{Shen}},\ and\ \bibinfo {author}
  {\bibfnamefont{N.~C.}\ \bibnamefont{Seeman}},\ }%
  \bibfield{journal}{%
  \bibinfo {journal} {J. Am. Chem. Soc.}\ }%
  \textbf{\bibinfo {volume} {124}},\ \bibinfo {pages} {12940} (\bibinfo {year}
  {2002})\BibitemShut{NoStop}%
\bibitem{Seeman01}%
  \BibitemOpen
  \bibfield{author}{%
  \bibinfo {author} {\bibfnamefont{N.~C.}\ \bibnamefont{Seeman}},\ }%
  \bibfield{journal}{%
  \bibinfo {journal} {Nano Letters}\ }%
  \textbf{\bibinfo {volume} {1}},\ \bibinfo {pages} {22} (\bibinfo {year}
  {2001})\BibitemShut{NoStop}%
\bibitem{Shen04}%
  \BibitemOpen
  \bibfield{author}{%
  \bibinfo {author} {\bibfnamefont{Z.}~\bibnamefont{Shen}}, \bibinfo {author}
  {\bibfnamefont{H.}~\bibnamefont{Yan}}, \bibinfo {author}
  {\bibfnamefont{T.}~\bibnamefont{Wang}},\ and\ \bibinfo {author}
  {\bibfnamefont{N.~C.}\ \bibnamefont{Seeman}},\ }%
  \bibfield{journal}{%
  \bibinfo {journal} {J. Am. Chem. Soc.}\ }%
  \textbf{\bibinfo {volume} {126}},\ \bibinfo {pages} {1666} (\bibinfo {year}
  {2004})\BibitemShut{NoStop}%
\bibitem{Liu08b}%
  \BibitemOpen
  \bibfield{author}{%
  \bibinfo {author} {\bibfnamefont{W.}~\bibnamefont{Liu}}, \bibinfo {author}
  {\bibfnamefont{X.}~\bibnamefont{Wang}}, \bibinfo {author}
  {\bibfnamefont{T.}~\bibnamefont{Wang}}, \bibinfo {author}
  {\bibfnamefont{R.}~\bibnamefont{Sha}},\ and\ \bibinfo {author}
  {\bibfnamefont{N.~C.}\ \bibnamefont{Seeman}},\ }%
  \bibfield{journal}{%
  \bibinfo {journal} {Nano Lett.}\ }%
  \textbf{\bibinfo {volume} {8}},\ \bibinfo {pages} {317} (\bibinfo {year}
  {2009})\BibitemShut{NoStop}%
\bibitem{Shih04}%
  \BibitemOpen
  \bibfield{author}{%
  \bibinfo {author} {\bibfnamefont{W.~M.}\ \bibnamefont{Shih}}, \bibinfo
  {author} {\bibfnamefont{J.~D.}\ \bibnamefont{Quispe}},\ and\ \bibinfo
  {author} {\bibfnamefont{G.~F.}\ \bibnamefont{Joyce}},\ }%
  \bibfield{journal}{%
  \bibinfo {journal} {Nature}\ }%
  \textbf{\bibinfo {volume} {427}},\ \bibinfo {pages} {618} (\bibinfo {year}
  {2004})\BibitemShut{NoStop}%
\bibitem{Wang10}%
  \BibitemOpen
  \bibfield{author}{%
  \bibinfo {author} {\bibfnamefont{X.}~\bibnamefont{Wang}}, \bibinfo {author}
  {\bibfnamefont{X.}~\bibnamefont{Zhang}}, \bibinfo {author}
  {\bibfnamefont{C.}~\bibnamefont{Mao}},\ and\ \bibinfo {author}
  {\bibfnamefont{N.~C.}\ \bibnamefont{Seeman}},\ }%
  \bibfield{journal}{%
  \bibinfo {journal} {Proc. Natl. Acad. Sci. USA}\ }%
  \textbf{\bibinfo {volume} {107}},\ \bibinfo {pages} {12547} (\bibinfo {year}
  {2010})\BibitemShut{NoStop}%
\bibitem{Kolb00}%
  \BibitemOpen
  \bibfield{author}{%
  \bibinfo {author} {\bibfnamefont{F.~A.}\ \bibnamefont{Kolb}}, \bibinfo
  {author} {\bibfnamefont{C.}~\bibnamefont{Malmgren}}, \bibinfo {author}
  {\bibfnamefont{E.}~\bibnamefont{Westhof}}, \bibinfo {author}
  {\bibfnamefont{C.}~\bibnamefont{Ehresmann}}, \bibinfo {author}
  {\bibfnamefont{B.}~\bibnamefont{Ehresmann}}, \bibinfo {author}
  {\bibfnamefont{E.~G.~H.}\ \bibnamefont{Wagner}},\ and\ \bibinfo {author}
  {\bibfnamefont{P.}~\bibnamefont{Romby}},\ }%
  \bibfield{journal}{%
  \bibinfo {journal} {RNA}\ }%
  \textbf{\bibinfo {volume} {6}},\ \bibinfo {pages} {311} (\bibinfo {year}
  {2000})\BibitemShut{NoStop}%
\bibitem{Kolb00b}%
  \BibitemOpen
  \bibfield{author}{%
  \bibinfo {author} {\bibfnamefont{F.~A.}\ \bibnamefont{Kolb}}, \bibinfo
  {author} {\bibfnamefont{H.~M.}\ \bibnamefont{Engdahl}}, \bibinfo {author}
  {\bibfnamefont{J.~G.}\ \bibnamefont{Slagter-J\"{a}ger}}, \bibinfo {author}
  {\bibfnamefont{B.}~\bibnamefont{Ehresmann}}, \bibinfo {author}
  {\bibfnamefont{C.}~\bibnamefont{Ehresmann}}, \bibinfo {author}
  {\bibfnamefont{E.}~\bibnamefont{Westhof}}, \bibinfo {author}
  {\bibfnamefont{E.~G.~H.}\ \bibnamefont{Wagner}},\ and\ \bibinfo {author}
  {\bibfnamefont{P.}~\bibnamefont{Romby}},\ }%
  \bibfield{journal}{%
  \bibinfo {journal} {EMBO J.}\ }%
  \textbf{\bibinfo {volume} {19}},\ \bibinfo {pages} {5905} (\bibinfo {year}
  {2000})\BibitemShut{NoStop}%
\bibitem{Kolb01}%
  \BibitemOpen
  \bibfield{author}{%
  \bibinfo {author} {\bibfnamefont{F.~A.}\ \bibnamefont{Kolb}}, \bibinfo
  {author} {\bibfnamefont{E.}~\bibnamefont{Westhof}}, \bibinfo {author}
  {\bibfnamefont{C.}~\bibnamefont{Ehresmann}}, \bibinfo {author}
  {\bibfnamefont{B.}~\bibnamefont{Ehresmann}}, \bibinfo {author}
  {\bibfnamefont{E.~G.~H.}\ \bibnamefont{Wagner}},\ and\ \bibinfo {author}
  {\bibfnamefont{P.}~\bibnamefont{Romby}},\ }%
  \bibfield{journal}{%
  \bibinfo {journal} {Nucl. Acids Res.}\ }%
  \textbf{\bibinfo {volume} {29}},\ \bibinfo {pages} {3145} (\bibinfo {year}
  {2001})\BibitemShut{NoStop}%
\bibitem{Kolb01b}%
  \BibitemOpen
  \bibfield{author}{%
  \bibinfo {author} {\bibfnamefont{F.~A.}\ \bibnamefont{Kolb}}, \bibinfo
  {author} {\bibfnamefont{E.}~\bibnamefont{Westhof}}, \bibinfo {author}
  {\bibfnamefont{B.}~\bibnamefont{Ehresmann}}, \bibinfo {author}
  {\bibfnamefont{C.}~\bibnamefont{Ehresmann}}, \bibinfo {author}
  {\bibfnamefont{E.~G.~H.}\ \bibnamefont{Wagner}},\ and\ \bibinfo {author}
  {\bibfnamefont{P.}~\bibnamefont{Romby}},\ }%
  \bibfield{journal}{%
  \bibinfo {journal} {J. Mol. Biol.}\ }%
  \textbf{\bibinfo {volume} {309}},\ \bibinfo {pages} {605} (\bibinfo {year}
  {2001})\BibitemShut{NoStop}%
\bibitem{Ouldridge_bulk_2010}%
  \BibitemOpen
  \bibfield{author}{%
  \bibinfo {author} {\bibfnamefont{T.~E.}\ \bibnamefont{Ouldridge}}, \bibinfo
  {author} {\bibfnamefont{A.~A.}\ \bibnamefont{Louis}},\ and\ \bibinfo {author}
  {\bibfnamefont{J.~P.~K.}\ \bibnamefont{Doye}},\ }%
  \bibfield{journal}{%
  \bibinfo {journal} {J. Phys.: Condens. Matter}\ }%
  \textbf{\bibinfo {volume} {22}},\ \bibinfo {pages} {104102} (\bibinfo {year}
  {2010})\BibitemShut{NoStop}%
\bibitem{Panyutin94}%
  \BibitemOpen
  \bibfield{author}{%
  \bibinfo {author} {\bibfnamefont{I.~G.}\ \bibnamefont{Panyutin}}\ and\
  \bibinfo {author} {\bibfnamefont{P.}~\bibnamefont{Hsieh}},\ }%
  \bibfield{journal}{%
  \bibinfo {journal} {Proc. Natl. Acad. Sci. USA}\ }%
  \textbf{\bibinfo {volume} {91}},\ \bibinfo {pages} {2021} (\bibinfo {year}
  {1994})\BibitemShut{NoStop}%
\bibitem{Rothemund06}%
  \BibitemOpen
  \bibfield{author}{%
  \bibinfo {author} {\bibfnamefont{P.~W.~K.}\ \bibnamefont{Rothemund}},\ }%
  \bibfield{journal}{%
  \bibinfo {journal} {Nature}\ }%
  \textbf{\bibinfo {volume} {440}},\ \bibinfo {pages} {297} (\bibinfo {year}
  {2006})\BibitemShut{NoStop}%
\end{thebibliography}

%Merlin.mbs v4.21 2009-07-09.
%

\end{document}